# Automated Identification of Thoracic Pathology from Chest Radiographs with Enhanced Training Pipeline

Adora M. DSouza[1], Anas Z. Abidin[2], and Axel Wismüller[1,2,3,4]
[1]Department of Electrical Engineering, University of Rochester, NY, USA
[2]Department of Biomedical Engineering, University of Rochester, NY, USA
[3]Department of Imaging Sciences, University of Rochester, NY, USA
[4]Faculty of Medicine and Institute of Clinical Radiology, Ludwig Maximilian University, Munich, Germany

## ABSTRACT

Chest x-rays are the most common radiology studies for diagnosing lung and heart disease. Hence, a system for automated pre-reporting of pathologic findings on chest x-rays would greatly enhance radiologists' productivity. To this end, we investigate a deep-learning framework with novel training schemes for classification of different thoracic pathology labels from chest x-rays. We use the currently largest publicly available annotated dataset ChestX-ray14 of 112,120 chest radiographs of 30,805 patients. Each image was annotated with either a 'NoFinding' class, or one or more of 14 thoracic pathology labels. Subjects can have multiple pathologies, resulting in a multi-class, multi-label problem. We encoded labels as binary vectors using $k$-hot encoding. We study the ResNet34 architecture, pre-trained on ImageNet, where two key modifications were incorporated into the training framework: (1) Stochastic gradient descent with momentum and with restarts using cosine annealing, (2) Variable image sizes for fine-tuning to prevent overfitting. Additionally, we use a heuristic algorithm to select a good learning rate. Learning with restarts was used to avoid local minima. Area Under receiver operating characteristics Curve (AUC) was used to quantitatively evaluate diagnostic quality. Our results are comparable to, or outperform the best results of current state-of-the-art methods with AUCs as follows: Atelectasis:0.81, Cardiomegaly:0.91, Consolidation:0.81, Edema:0.92, Effusion:0.89, Emphysema: 0.92, Fibrosis:0.81, Hernia:0.84, Infiltration:0.73, Mass:0.85, Nodule:0.76, Pleural Thickening:0.81, Pneumonia:0.77, Pneumothorax:0.89 and NoFinding:0.79. Our results suggest that, in addition to using sophisticated network architectures, a good learning rate, scheduler and a robust optimizer can boost performance.

**Keywords:** Chest x-ray, computer aided diagnosis, deep learning, transfer learning, fine-tuning, ResNet34

## 1. INTRODUCTION

One of the key challenges in the medical image analysis domain is to assist radiologists in analyzing the growing number of imaging studies per time. Computer aided diagnosis (CAD) is a step towards reducing the burden on a radiologist by providing meaningful aid, complementary to the radiologist's work. However, growth and advancement of CAD techniques requires vast amounts of curated data. Initiatives to make anonymized data available to the scientific community can expedite the development of sophisticated algorithms for disease identification, segmentation, and other image analysis tasks.

---

[1] adora.dsouza@rochester.edu; University of Rochester, NY

The recent release of the largest publicly available chest x-ray dataset with disease labels by Wang et al. [1] is a step towards this aim. Chest x-rays are commonly used for diagnosing and screening of a multitude of lung diseases. Wang et al. [1] demonstrated with promising results that the commonly occurring thoracic disease labels are detectable with deep convolutional neural network learning. An automated detection of the most common thoracic disease labels from chest x-rays will decrease the burden on radiologists, and potentially also make healthcare accessible to communities with shortage of radiologists.

In this work, we demonstrate how a simple network architecture, such as the ResNet34 [2], per-trained on ImageNet, fine-tuned, while adopting practices to learn at a good rate without compromising on accuracy and reducing overfitting, can achieve performance similar to or better than some of the more sophisticated architectures. In our study, learning rate is selected using a heuristic algorithm [3]. In addition, we use stochastic gradient decent with restarts and cosine annealing for learning rate scheduling [4]. Furthermore, to improve generalizability, fine tuning is carried out with increasing image sizes. We also conduct an ablation study for investigating the different components of the training procedure.

## 2. DATA

**Chest X-Ray data**

We use frontal-view chest x-rays of the currently largest publicly available annotated dataset [1] of 112,120 radiographs of 30,805 patients. Each image was annotated with either a 'no finding' class, or one or more of 14 common thoracic pathology labels. The authors obtained disease labels associated with each image by mining radiological text reports using natural language processing techniques, details can be found in [1]. All images were resized to 1024 x 1024 without preserving aspect ratio. In this work, we divide the data into train (70% of subjects), validation (10% of subjects) and test (20% of subjects) sets. We ensured strict separation of the data with no overlap of individual subject's scans between splits.

## 3. METHODS

In this work, we explore if fine-tuning a simple architecture for performing multiclass multi-label classification can be improved with good learning rate schedulers, and techniques to avoid overfitting and improve generalizability. We adopt the ResNet34 architecture [2] pre-trained on ImageNet.

**Residual Networks:**

Residual networks were introduced in [2] as a network to solve the vanishing gradients problem by introducing shortcuts or skip connections. These connections make sure that the gradient signal propagate back without getting diminished. In traditional neural networks, the output can be defined as $y = f(x)$ where $f$ is a convolution, batch normalization etc. on the input $x$, resulting in output $y$. In such a network the gradient has to pass through $f(x)$ during back propogation which can cause issues when nonlinearities are involved. The residual network provides an elegant solution to this problem by introducing a shortcut for the gradient to pass through: $y = f(x) + x$. We adopt the 34-layer residual network, called ResNet34.

**Multi-label formulation:**

We use $k$-hot encoded vector for the binary multi-class multi-label problem, i.e. the output $\mathbf{y}_{disease\_state} = [y_1, y_2, \ldots y_c, .. y_C]$ where, $y_c \in \{0, 1\}$ and $C = 15$. The first 14 of the outputs correspond to the 14 disease states, while the 15th output corresponds to the absence of any disease (No Finding). Previous work has used $C = 14$ with absence of disease being all 0s i.e. $\mathbf{y}_{No\_finding} = [0, 0, \ldots, 0]$. Our motivation for including a 15th class was for the ease of calculation of performance on healthy subjects as well.

**Network architecture and training:**

We use the negative log likelihood loss function, which is useful for training a multiclass classification problem. ResNet34 [2] pre-trained on ImageNet was used. Data was divided into batches of size = 50, and the maximum size of the scans used was 340 x 340. One of the key parameters to be set before training any deep learning algorithm is the learning rate. In [3] the authors provide a heuristic approach to find a good learning rate prior to training, which we incorporated in our analysis. Additionally, we use stochastic gradient descent with momentum (SGDM) as the optimizer. The learning rate is varied using cosine annealing with restarts [4], reducing chances of the cost function optimization being trapped in a local minimum. To avoid overfitting, we fine-tune the network in steps of increasing data dimension. Initially, the network is fine-tuned on data of dimension 64 x 64, followed by 128 x 128, then 256 x 256, and finally 340 x 340. The training pipeline is provided in Figure 1.

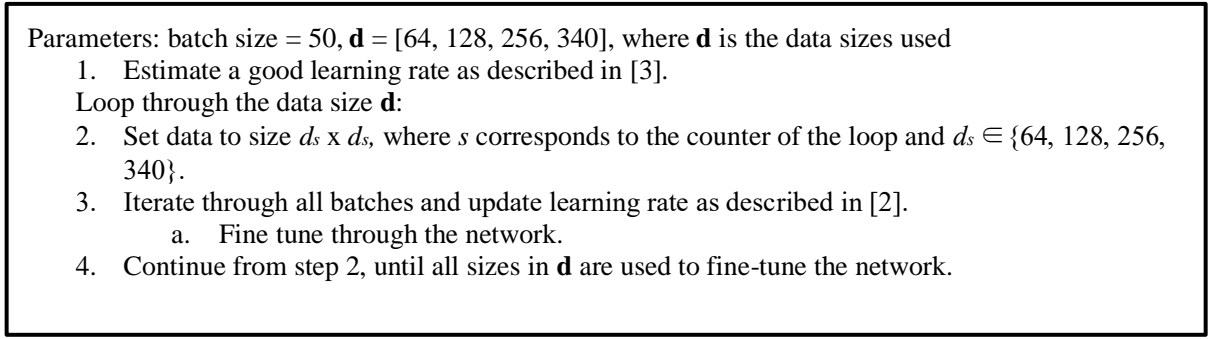

Parameters: batch size = 50, **d** = [64, 128, 256, 340], where **d** is the data sizes used
1. Estimate a good learning rate as described in [3].
Loop through the data size **d**:
2. Set data to size $d_s$ x $d_s$, where $s$ corresponds to the counter of the loop and $d_s \in \{64, 128, 256, 340\}$.
3. Iterate through all batches and update learning rate as described in [2].
    a. Fine tune through the network.
4. Continue from step 2, until all sizes in **d** are used to fine-tune the network.

**Figure 1**: Training pipeline

The above techniques for avoiding overfitting of the data were adopted from the publicly available fast.ai library (http://www.fast.ai/) [5]. Area Under receiver operating characteristics Curve (AUC) was used to quantitatively evaluate diagnostic quality.

**Ablation study:**

We also study effects of the various components in the above pipeline by carrying out an ablation study as follows:

(i) Fine-tuning is carried out only on one size, i.e. 340 x 340 (**d** = [340]), keeping all the other components the same; this training pipeline is referred to as V1.
(ii) Here, **d** is left unchanged, i.e. **d** = [64, 128, 256, 340]. However, optimization using SDGM without restarts in the absence of cosine annealing is carried out (V2).
(ii) Finally, fine-tuning is carried out only on one size, i.e. 340 x 340 (**d** = [340]), and SGDM optimization without restarts in the absence of cosine annealing is used.

Additionally, we compare performance with two state-of-the-art approaches, namely one developed by Li et al [6], which we refer to as S1, and a second one from CheXNet [7], which we refer to as S2. An overall training time of 3.5 hours, with an NVIDIA Tesla P100 GPU, was needed for the analysis.

All implementations were carried out using the fast.ai library [5], which is built on PyTorch.

## 4. RESULTS

Results of the ablation study are shown in Table 1. Here, we clearly see that including fine tuning for different image scales and scheduling the learning rate using cosine annealing with restarts for SGDM boosts overall performance. Another interesting finding is that of the three ablation studies, V1, V2 and V3, V2 performs the best. This indicates that repeated fine-tuning of the network with different data sizes boosts performance. This intuitively makes sense since the network is presented with different sizes of a pathology, making it scale-invariant.

Table 2 compares our approach with current state of the art methods. Our approach performs comparably well as CheXNet. Additionally, it outperforms the results from Li et al [6] in predicting all pathologies.

**Table 1**: AUC results of our ablation study, for the different thoracic pathologies. The proposed approach works best justifying that performance degrades upon removal of any of the components included in the training pipeline. V1: Fine-tuning size 340 x 340 alone, V2: Optimization without restarts in the absence of cosine annealing, V3: Fine-tuning size 340 x 340 alone and optimization without restarts in the absence of cosine annealing.

| Pathology | Proposed | V1 | V2 | V3 |
|---|---|---|---|---|
| Atelectasis | **0.8143** | 0.7975 | 0.8014 | 0.7989 |
| Cardiomegaly | **0.9129** | 0.8989 | 0.9026 | 0.8969 |
| Consolidation | 0.811 | 0.8031 | **0.8126** | 0.7997 |
| Edema | **0.922** | 0.9099 | 0.9163 | 0.9077 |
| Effusion | **0.8884** | 0.8745 | 0.8829 | 0.8777 |
| Emphysema | **0.9174** | 0.9017 | 0.9083 | 0.9046 |
| Fibrosis | **0.8148** | 0.7918 | 0.8055 | 0.7992 |
| Hernia | 0.8388 | 0.7998 | **0.8525** | 0.8268 |
| Infiltration | **0.7265** | 0.7216 | 0.7208 | 0.7162 |
| Mass | **0.8487** | 0.8153 | 0.8318 | 0.8123 |
| No Finding | **0.7889** | 0.7767 | 0.782 | 0.7793 |
| Nodule | **0.7553** | 0.7253 | 0.7296 | 0.7166 |
| Pleural Thickening | **0.8076** | 0.796 | 0.793 | 0.7899 |
| Pneumonia | 0.7698 | 0.7685 | **0.7741** | 0.7617 |
| Pneumothorax | **0.8884** | 0.8779 | 0.8839 | 0.8693 |

## 5. DISCUSSION

Our results demonstrate that a simple network structure can produce good results given that the learning procedure and loss functions are selected well. In this work, we encode thoracic pathology labels as binary vectors using *k*-hot coding for each of the classes. We fine-tune the 34 layer ResNet-34 architecture [2], pre-trained on ImageNet.

The two key modifications we explore for improving network performance are: (1) SGDM with restarts using cosine annealing, (2) Variable image sizes for training to prevent overfitting. We explore these two key modifications with an ablation study. The results of this ablation study (Table 1) demonstrate that these two modifications are useful in improving performance. Furthermore, comparing this method with other approaches in literature, we achieve comparable performance. In the future, we aim to divide the dataset into a number of different train-test pairs and obtain a measure of average performance and the distribution of accuracy across different pairs.

Table 2: AUC results of our study, compared to other approaches in literature, for classifying the different thoracic pathologies. The proposed approach is comparable to CheXNet [7], i.e. S2 and is better than S1 [6] in detecting all pathologies.

| Pathology | S1 [6] | S2 [7] | Proposed |
| --- | --- | --- | --- |
| Atelectasis | 0.80 ± 0.00 | 0.8094 | **0.8143** |
| Cardiomegaly | 0.87 ± 0:01 | **0.9248** | 0.9129 |
| Consolidation | 0.80 ± 0:01 | 0.7901 | **0.811** |
| Edema | 0.88 ± 0:01 | 0.8878 | **0.922** |
| Effusion | 0.87 ± 0:00 | 0.8638 | **0.8884** |
| Emphysema | 0.91 ± 0:01 | **0.9371** | 0.9174 |
| Fibrosis | 0:78 ± 0:02 | 0.8047 | **0.8148** |
| Hernia | 0:77 ± 0:03 | **0.9164** | 0.8388 |
| Infiltration | 0.70 ± 0:01 | **0.7345** | 0.7265 |
| Mass | 0.83 ± 0:01 | **0.8676** | 0.8487 |
| No Finding | - | - | 0.7889 |
| Nodule | 0.75 ± 0:01 | **0.7802** | 0.7553 |
| Pleural Thickening | 0.79 ± 0:01 | 0.8062 | **0.8076** |
| Pneumonia | 0.67 ± 0:01 | 0.7680 | **0.7698** |
| Pneumothorax | 0.87 ± 0:01 | **0.8887** | 0.8884 |

## 6. CONCLUSION

We present an approach to effectively classify thoracic pathology labels from chest x-ray data. To this end, we adopt a number of steps to improve learning on the 34 layer ResNet architecture [2]. The Chest x-ray14 data was used in this analysis and a strict data separation between the train, validation and test groups were carried out, also ensuring that multiple scans belonging to an individual were kept in either one group. For this multi-class multi-label classification problem, the negative log likelihood loss function was used. Additionally, we adopted a heuristic approach to estimate a good learning rate [3], optimized with SGDM with restarts using cosine annealing [4], and used variable image sizes for training to prevent overfitting. Our ablation study supports the usefulness of the various components to improve training of the network. These training steps appear to boost performance considerably, and we have achieved comparable performance to state of the art methods. Our results suggest that, in addition to using sophisticated network architectures, choosing a good learning rate, scheduler and a robust optimizer can boost performance of the classifier. We can thus conclude that the training approaches presented in our work should be adopted for better training and improved generalizability.

## 7. ACKNOWLEDGEMENTS

This research was funded by the National Institutes of Health (NIH) Award R01-DA-034977. The content is solely the responsibility of the authors and does not necessarily represent the official views of the National Institute of Health. This work was conducted as a Practice Quality Improvement (PQI) project related to American Board of Radiology (ABR) Maintenance of Certificate (MOC) for Prof. Dr. Axel Wismüller. The authors would like to thank the founders of fast.ai (http://www.fast.ai), Dr. Jeremy Howard and Dr. Rachel Thomas.

This work is not being and has not been submitted for publication or presentation elsewhere.

# REFERENCES

[1] Wang, X., Peng, Y., Lu, L., Lu, Z., Bagheri, M., and Summers, R. M., "Chestx-ray8: Hospital-scale chest x-ray database and benchmarks on weakly-supervised classification and localization of common thorax diseases," In Computer Vision and Pattern Recognition (CVPR), 2017 IEEE Conference on, 3462-3471 (2017)

[2] He, K., Zhang, X., Ren, S., and Sun, J., "Deep residual learning for image recognition," In Proceedings of the IEEE conference on computer vision and pattern recognition, 770-778 (2016)

[3] Smith, L. N., "Cyclical learning rates for training neural networks," In Applications of Computer Vision (WACV), 2017 IEEE Winter Conference on, 464-472 (2017)

[4] Loshchilov, I., and Hutter, F., "Sgdr: Stochastic gradient descent with warm restarts," arXiv preprint arXiv:1608.03983 (2016).

[5] Fast.ai, Making neural nets uncool again: http://www.fast.ai/

[6] Li, Z., Wang, C., Han, M., Xue, Y., Wei, W., Li, L.J. and Fei-Fei, L., "Thoracic disease identification and localization with limited supervision," In *Proceedings of the IEEE Conference on Computer Vision and Pattern Recognition* 8290-8299 (2018).

[7] Rajpurkar, P., Irvin, J., Zhu, K., Yang, B., Mehta, H., Duan, T., Ding, D., Bagul, A., Langlotz, C., Shpanskaya, K., Lungren, M.P. and Ng, A.Y., "Chexnet: Radiologist-level pneumonia detection on chest x-rays with deep learning," arXiv preprint arXiv:1711.05225 (2017).